\documentstyle[psfig]{elsart}

\begin{document}

\begin{frontmatter}

\title{Triplet pairing of neutrons in $\beta$-stable neutron star matter}

\author[oslo]{\O.\ Elgar\o y,}
\author[oslo]{L.\ Engvik,}
\author[ect]{M.\ Hjorth-Jensen,}
\author[oslo]{E.\ Osnes}
\address[oslo]{Department of Physics, 
        University of Oslo, N-0316 Oslo, Norway}
\address[ect]{ECT*, European Centre for Theoretical Studies in 
        Nuclear Physics and Related Areas, 
        Strada delle Tabarelle 286, I-38050 Villazzano (Trento), Italy}

\maketitle

\begin{abstract}
$^{3}P_{2}$ pairing in neutron matter is investigated using the Bonn 
potential models.  We find pairing energy gaps in pure 
neutron matter comparable to the results of previous investigators 
when the attractive tensor coupling is included.  
However, taking into account that in a neutron star we have matter at 
$\beta$ equilibrium, we find that the 
$^{3}P_{2}$-$^{3}F_{2}$ energy gap is reduced considerably. 
\end{abstract}

\end{frontmatter}

\section{Introduction}

Superfluidity in neutron stars is expected to have a profound influence 
on their dynamical and thermal properties \cite{sha83}.  
Due to the properties of nuclear forces several nucleon superfluids can 
be formed, the most commonly considered being the $^{1}S_{0}$  
neutron and proton superfluids, and the $^{3}P_{2}$ neutron 
superfluid.  In a recent investigation \cite{elg96} we calculated   
$^{1}S_{0}$ neutron and proton pairing gaps, ignoring complicated 
many-body effects.  In this paper we follow the same philosophy  
in calculating energy gaps for neutrons in the $^{3}P_{2}$ 
state, using realistic nucleon-nucleon (NN) forces derived from 
meson theory \cite{mac89}.

\section{Theory}

The standard BCS theory \cite{bcs57,cms59} has to be extended when dealing 
with pairing of particles in states with non-zero angular momentum.  
The method 
of generalizing the Bogolyubov transformation to pairing of particles 
with arbitrary angular momentum and isospin is described in 
refs. \cite{tam70,tak93}.  We are in 
this paper concerned with neutron pairing in the $^{3}P_{2}$-$^{3}F_{2}$ 
state, and thus take $J=2$, $S=1$, $L=1,3$, and $T=1$.  In this case, the 
$^{1}S_{0}$ 
gap equation is replaced by a set of ten ($2\times (2J+1)$) coupled equations 
for the ten energy gap components:
\begin{eqnarray}
\Delta_{\lambda L m_{J}}(k)&=&-\frac{1}{\pi}(-1)^{1-S}\int_{0}^{\infty}
dk'k'^{2}\sum_{L'}i^{L'-L}\langle k'|V_{\lambda}^{L'L}|k\rangle \nonumber \\ 
&\times&\int d{\bf \hat{k'}}\sum_{L''m_{J}'}\Delta_{\lambda L''m_{J}'}(k')
{\rm Tr}[{\bf G}^{*}_{\lambda L'm_{J}}({\bf \hat{k'}}){\bf G}_{\lambda 
L'' m_{J}'}(\bf{\hat{k'}})] \nonumber \\
&\times&\frac{1}{\sqrt{(\epsilon_{k'}-\epsilon_{k_{F}})^{2}+D_{\lambda}
^{2}({\bf k'})}} 
\label{eq:tgap1} 
\end{eqnarray}  
where $\lambda\equiv(T=1,S=1,J=2)$, the orbital angular momenta $L,\;L',$
and $L''$ take on the values $1\;{\rm and}\;3$, and 
\[
{\bf G}_{\lambda L m_{J}}({\bf \hat{k}})\equiv \left \{ \left \langle 
\frac{1}{2}\sigma_{1}\frac{1}{2}\sigma_{2}|Sm_{S} \right \rangle 
\left \langle S m_{S} L m_{L} | J m_{J} \right \rangle Y_{L m_{J}-m_{S}}
({\bf \hat{k}}) \right \}
\]
is a $2\times 2$ matrix in spin-$\frac{1}{2}$ space.  Furthermore,  
$Y_{L m_{J}-m_{S}}$ is a spherical harmonic, $D_{\lambda}$ is 
given by 
\[
D_{\lambda}^{2}({\bf k})=\frac{1}{2}\sum_{LL'}\sum_{m_{J}m_{J}'}
\Delta^{*}_{\lambda L m_{J}}(k)\Delta_{\lambda L m_{J}'}(k) 
{\rm Tr}[{\bf G}^{\dagger}_{\lambda L m_{J}}({\bf \hat{k}})
{\bf G}_{\lambda L' m_{J}'}({\bf \hat{k}})], 
\]
and $\epsilon_{k}$ and $\epsilon_{k_{F}}$ are single-particles, evaluated at 
$k=k$ and $k=k_{F}$, respectively.  The symbol ${\rm Tr}$ denotes 
the trace operation, the symbol $^{\dagger}$ denotes Hermitian conjugation, 
and $\left \langle k' |V_{\lambda}^{LL'} |k \right \rangle$ 
are matrix elements of the NN interaction in momentum space,
 in the partial wave characterized by $L,L'$ and $\lambda$.

\subsection{Uncoupled equations for $^{3}P_{2}$ pairing}
The set of equations (\ref{eq:tgap1}) is rarely solved in its full 
generality, and we will in this paper look at two approximations.  
In the first one, considered by e.g. Baldo {\it et al.} \cite{bld92},   
the tensor interaction is neglected, and thus pairing in a pure 
$^{3}P_{2}$ wave is considered.
Furthermore, only the solutions with $m_{J}=0$ or $|m_{J}|=2$ are considered, 
because these have been found to be close to the most general solution of 
eq. (\ref{eq:tgap1}) \cite{tak93}.   
It can then be shown that the gap equations reduce to 
\begin{equation}
\Delta_{i}(k)=-\frac{1}{\pi}\int_{0}^{\infty}dk'k'^{2} \left \langle k' |
V(^{3}P_{2}) |k \right \rangle \frac{\Delta_{i}(k')}{\tilde{E}_{i}
(k')}, \;i=0,2, 
\label{eq:tripg}
\end{equation}
where $i=0,2$ for $m_{J}=0,\; |m_{J}|=2$ respectively.  We have here 
defined 
\begin{equation}
   \frac{1}{\tilde{E}_{i}(k)}\equiv
    \left\{\begin{array}{cc}\frac{3}{4}\int_{-1}^{+1}d(\cos \theta)
    \frac{1-\cos^{2}\theta}{E_{2}({\bf k})},&i=2,\\
    \frac{1}{4}\int_{-1}^{+1}d(\cos \theta)\frac{1+3\cos^{2}\theta}
    {E_{0}({\bf k})},&i=0,\end{array}\right.
\end{equation}
where $E({\bf k})$ is the quasiparticle energy, given by 
\begin{equation}
   E_{i}({\bf k})=
   \left\{\begin{array}{cc}\sqrt{(\epsilon_{k}-\epsilon_{k_{F}})^{2}
    +\frac{3}{8\pi}\Delta_{2}^{2}(1-\cos^{2}\theta)},&i=2,\\ 
   \sqrt{(\epsilon_{k}-\epsilon_{k_{F}})^{2}+\frac{1}{16\pi}\Delta_{0}
   ^{2}(k)(1+3\cos^{2}\theta)},&i=0.\end{array}\right.
\label{eq:qua1}
\end{equation}
In eq. (\ref{eq:tripg}) $\left \langle k' |V(^{3}P_{2}) | k \right \rangle$ 
are the momentum-space matrix elements of the bare NN interaction 
in the $^{3}P_{2}$ wave.   
Eq. (\ref{eq:tripg}) must in each case be solved self-consistently 
for the gap component $\Delta_{0}(k)$ or $\Delta_{2}(k)$.  
The energy gap is by convention defined to be the second term appearing 
under the square root in the expression for the quasiparticle energy:
\begin{equation}
D_{0}^{2}({\bf k})=\frac{1}{16\pi}\Delta_{0}^{2}(k)(1+3\cos^{2}\theta)
\label{eq:d0a}
\end{equation}
\begin{equation}
D_{2}^{2}({\bf k})=\frac{3}{8\pi}\Delta_{2}^{2}(k)(1-\cos^{2}\theta) 
\label{eq:d2a}
\end{equation}  
When giving numerical results, we use angle-averaged energy gaps, given 
by 
\begin{eqnarray}
\overline{D_{2}^{2}(k)}=\frac{1}{4\pi}\Delta_{2}^{2}(k) \nonumber \\
\overline{D_{0}^{2}(k)}=\frac{1}{8\pi}\Delta_{0}^{2}(k). \nonumber 
\end{eqnarray}
Replacing angle-dependent quantities in the gap equation with their 
angular average has been found to be a good approximation \cite{bld92}.  
However, the angle dependence is easily handled numerically, so we have 
throughout solved the gap equation with angle dependencies included.  
 
\subsection{Coupled equations for $^{3}P_{2}$-$^{3}F_{2}$ pairing}

In the second approximation we include the tensor coupling effect, but 
look at the solution with $m_{J}=0$ only, thus reducing the number 
of components from ten to two.  This solution is expected to be a 
good approximation to the most general solution of eq. (\ref{eq:tgap1}) 
\cite{tak93}.
For $m_{J}=0$, the two coupled equations 
for the $L=1$ and $L=3$ gap components become
\begin{eqnarray}
\Delta_{\lambda 10}(k)&=&-\frac{1}{\pi}\int_{0}^{\infty}dk'k'^{2}
\left \langle k'| V_{\lambda}^{11} |k \right \rangle \int d{\bf \hat{k'}}
\frac{\left \{ \Delta_{\lambda 10}(k')f(\theta)+\Delta_{\lambda 30}(k')
g(\theta) \right \}}{E({\bf k'})} \nonumber \\
&+&\frac{1}{\pi}\int_{0}^{\infty}dk'k'^{2}\left \langle k'|V_{\lambda}^{31}|
k \right \rangle \int d{\bf \hat{k'}}\frac{\left \{\Delta_{\lambda 10}(k')
g(\theta) +\Delta_{\lambda 30}(k')h(\theta)\right \}}{E({\bf k'})}  
\label{eq:tgap2}
\end{eqnarray}
\begin{eqnarray}
\Delta_{\lambda 30}(k)&=&\frac{1}{\pi}\int_{0}^{\infty}dk'k'^{2} 
\left \langle k'|V_{\lambda}^{13}|k\right \rangle\int d{\bf \hat{k'}} 
\frac{\left \{\Delta_{\lambda 10}(k')f(\theta)+\Delta_{\lambda 30}(k')
g(\theta) \right \}}{E({\bf k'})} \nonumber \\
&-&\frac{1}{\pi}\int dk'k'^{2}\left \langle k'|V_{\lambda}^{33}|k\right\rangle
 \int d{\bf \hat{k'}}\frac{\left \{\Delta_{\lambda 10}(k')g(\theta)
+\Delta_{\lambda 30}(k')h(\theta) \right \}}{E({\bf k'})}. 
\label{eq:tgap3}
\end{eqnarray}  
The functions of the polar angle $\theta$ of ${\bf k'}$ are defined by  
\begin{eqnarray}
f(\theta)&=&\frac{1}{8\pi}\left(1+3\cos^{2}\theta \right ), \nonumber \\
g(\theta)&=&-\frac{\sqrt{6}}{64\pi}\left (1-7\cos^{2}\theta+5\sin\theta
\sin 3\theta-10\cos\theta\cos 3\theta \right ) \nonumber \\
h(\theta)&=&\frac{3}{128\pi}\left (13+4\cos^{2}\theta +5\sin\theta\sin 3
\theta+15\cos\theta\cos 3\theta \right ), \nonumber
\end{eqnarray}
the quasiparticle energy $E({\bf k})$ is given by 
\begin{equation}
E({\bf k})=\sqrt{(\epsilon_{k}-\epsilon_{k_{F}})^{2}+D_{\lambda}^{2}({\bf k})
},  
\label{eq:quasie}
\end{equation}
and the energy gap by 
\begin{equation}
D_{\lambda}^{2}({\bf k})=\frac{1}{2}f(\theta)\Delta_{\lambda 10}^{2}(k)
+\frac{1}{2}h(\theta)\Delta_{\lambda 30}^{2}(k)+g(\theta)\Delta_{\lambda 10}
(k)\Delta_{\lambda 30}(k).
\label{eq:dfunc}
\end{equation}
In ref. \cite{bld92} Baldo {\it et al.} obtained a large 
$^{3}P_{2}$ energy gap 
without the tensor coupling.  However, in refs. \cite{tak93,am85} the 
added attraction from the tensor coupling effect was found to be 
essential to the existence of superfluidity in this state.  
The importance of the tensor coupling depends to some extent on the 
choice of NN interaction, and we will therefore consider 
both cases.

\subsection{Calculation of single-particle energies}
The single-particle energies appearing in the quasiparticle energies  
(\ref{eq:qua1}) and (\ref{eq:quasie}) are obtained through a self-consistent 
BHF calculation, 
using a $G$-matrix defined through the Bethe-Brueckner-Goldstone
equation as  
\begin{equation}
   G=V+V\frac{Q}{\omega -H_0}G,
\end{equation}
where $V$ is the nucleon-nucleon potential, $Q$ is the Pauli operator
which prevents scattering into intermediate 
states prohibited by the Pauli
principle, $H_0$ is the unperturbed 
Hamiltonian acting on the intermediate
states and $\omega$ is the so-called starting energy, 
the unperturbed energy
of the interacting particles. Methods to solve this equation are reviewed in
ref. \cite{hko95}.
The single-particle energy for state $k_i$ ($i$ 
encompasses all relevant
quantum numbers like momentum, 
isospin projection, spin etc.)
in nuclear matter is assumed to have
the simple quadratic form\footnote{Throughout this work we set $\hbar=c=1$.}
\begin{equation}
   \epsilon_{k_i}=
   {\displaystyle\frac{k_{i}^2}
   {2m^{*}}}+\delta_i ,
   \label{eq:spen}
\end{equation}
where $m^{*}$ is the effective mass.
The terms $m^{*}$ and $\delta$, the latter being 
an effective single-particle
potential related to the $G$-matrix, are obtained through the
self-consistent BHF  procedure. The so-called
model-space BHF method 
for the single-particle spectrum has been used, see
e.g. ref. \cite{hko95}, with a cutoff momentum  
$k_M=3.0$ fm$^{-1}>k_{F}$.
In this approach the single-particle spectrum is defined by 
\begin{equation}
\epsilon_{k_{i}}=\frac{k_{i}^{2}}{2m}+u_{i}, 
\label{eq:modsp}
\end{equation}
where m is the nucleon mass, and the single-particle potential $u_{i}$ is 
given by 
\begin{equation}
    u_{i}=
    \left\{\begin{array}{cc}\sum_{k_{h}\leq k_{F}}\left\langle 
 k_{i}k_{h}|G(\omega=\epsilon_{k_{i}}+\epsilon_{k_{h}})|k_{i}k_{h}
\right\rangle,&k_{i}\leq k_{M},\\
    0,k_{i}>k_{M},\end{array}\right.,
\label{eq:modpot}
\end{equation}
where the subscript $AS$ denotes antisymmetrized matrix elements.  
This prescription reduces the discontinuity in the single-particle spectrum 
as compared with the standard BHF choice $k_{M}=k_{F}$.  
The self-consistency scheme 
consists in choosing adequate initial values of the
effective mass and $\delta$. The obtained $G$-matrix is in then used to 
calculate the single-particle potential $u_{i}$, from which we obtain 
new values for $m^{*}$ and $\delta$.  
This procedure continues until these parameters vary little. 
We stress that in the energy gap calculations we have used the 
single-particle spectrum defined by eqs. (\ref{eq:modsp}) and 
(\ref{eq:modpot}), and not the effective mass approximation (\ref{eq:spen}).

The BHF equations can be solved
for different proton fractions, using the formalism of refs. 
\cite{hko95,ks93}. The conditions for $\beta$ equilibrium require
that 
\begin{equation}
     \mu_{n}=\mu_{p}+\mu_{e},
\end{equation}
where $\mu_i$ is the chemical potential of particle type $i$, 
and that charge is conserved
\begin{equation}
     n_{p}=n_{e},
\end{equation}
where $n_{i}$ is the particle number density for particle $i$.  If 
muons are present, the condition for charge conservation becomes 
\begin{equation}
n_{p}=n_{e}+n_{\mu},
\label{eq:chcon2}
\end{equation}
and conservation of energy requires that 
\begin{equation}
\mu_{e}=\mu_{\mu}.
\end{equation}
We assume that neutrinos escape freely from the neutron star.  
The proton and neutron chemical potentials are determined from the 
energy per baryon, calculated self-consistently in the MBHF approach.  
The electron chemical potential, and thereby the muon chemical potential, 
is then given by $\mu_{e}=\mu_{n}-\mu_{p}$.  The Fermi momentum of lepton 
type $l=e,\mu$ is found from 
\begin{equation}
k_{F_{l}}=\mu_{l}^{2}-m_{l}^{2}
\end{equation}
where $m_{l}$ is the mass of lepton $l$, and we get the particle density 
using $n_{l}=k_{l}^{3}/3\pi^{2}$.  The proton fraction is then determined 
by the charge neutrality condition (\ref{eq:chcon2}).
 
Throughout this work, we use the meson-exchange potential models of the 
Bonn group, versions A, B and C in Table A.1 of ref. \cite{mac89}.  
These potentials 
have given satisfactory results in applications to finite nuclei and 
nuclear matter \cite{hko95}.  Furthermore, the only parameters 
entering in the Bonn models are physically motivated ones, e.g. meson 
masses.  A point of interest in this work is that the Bonn potentials 
have a rather weak tensor component.  
The A.1 parameters are the consistent choice of parameters in 
a non-relativistic approach, as they were 
fitted using the Blankenbecler-Sugar reduction \cite{BbS66} of the 
four-dimensional 
Bethe-Salpeter equation, yielding a 
$T$-matrix equation of the same form as the non-relativistic 
Lippmann-Schwinger equation \cite{mac89}.       

\section{Pairing in the pure $^{3}P_{2}$ wave}

\begin{figure}[htbp]
\begin{center}
{\centering
\mbox
{\psfig{figure=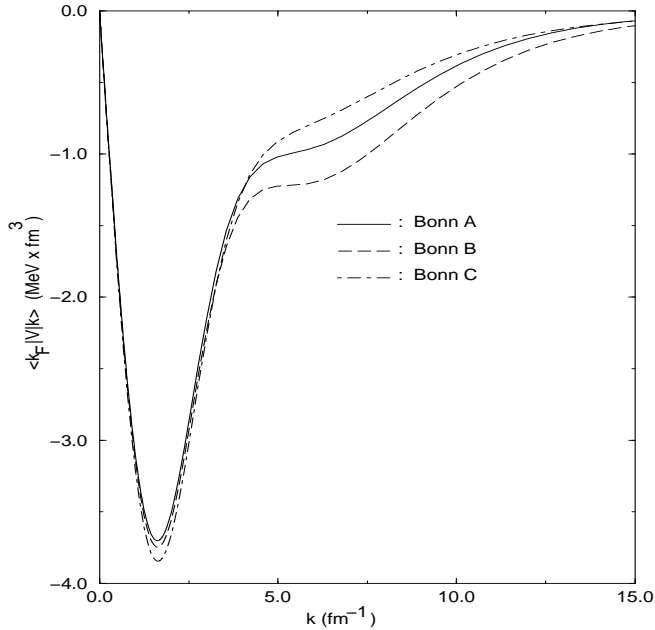,height=8cm,width=8cm}}
}
\caption{Matrix elements $\left \langle k_{F}|V(^{3}P_{2})|k \right \rangle$ 
of the Bonn A, B and C potentials, shown as 
 functions of the relative momentum $k$ for $k_{F}=1.8\;{\rm fm}^{-1}$. 
}
\label{fig:fig1}
\end{center}
\end{figure}

\begin{figure}[htbp]
\begin{center}
{\centering
\mbox
{\psfig{figure=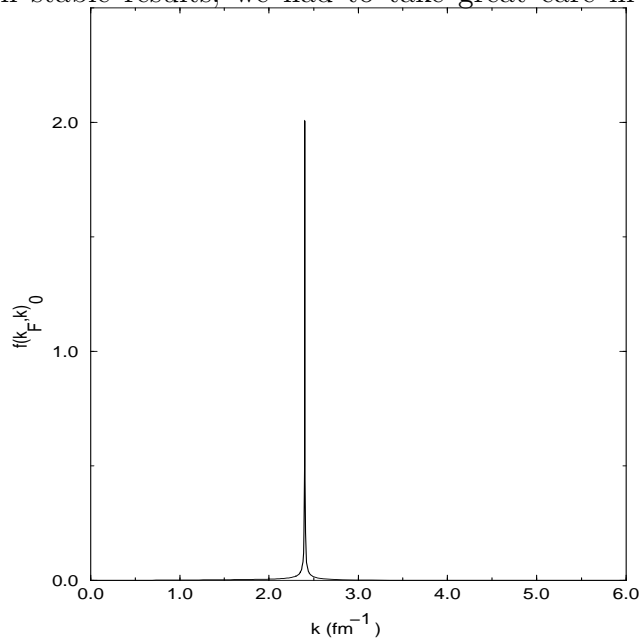,height=8cm,width=8cm}}
}
\caption{Integrand $f(k,k')_{0}$ in the gap equation for $m_{J}=0$ and 
$k=k_{F}=2.4\;{\rm fm}^{-1}$.    
}
\label{fig:fig2}
\end{center}
\end{figure}

We have solved the gap equations (\ref{eq:tripg}) in pure neutron matter.  
The calculational procedure consists of two steps:  First we obtain 
self-consistent single-particle energies in the MBHF approach, as described 
in section 2.3.  The next step is to solve the relevant gap equation, 
for $m_{J}=0$ or $|m_{J}|=2$, with the bare NN interaction as pairing 
interaction.  Being non-linear integral equations, the gap equations 
must be solved by iteration.  In ref. \cite{elg96} we applied a 
model-space approach to the gap equation for $^{1}S_{0}$ pairing.  
One of the advantages of this approach was that the numerical treatment 
became easier, as it allowed us to treat pairing correlations and 
the effects of the repulsive core of the NN interaction separately. 
The same method can easily be applied to the gap equations studied in this 
paper.  However, as seen from fig. \ref{fig:fig1}, the interaction 
in the $^{3}P_{2}$ wave is weak at 
the densities of interest, so a model space treatment is not called for. 
Nevertheless, we would like to point out some  
problems encountered in the numerical solution.  
First, the $^{3}P_{2}$ energy gap is considerably smaller than the 
$^{1}S_{0}$ energy gap, which is of order 1 MeV \cite{elg96}.  
A small energy gap at the Fermi surface causes problems in the   
numerical treatment, as we see from eq. (\ref{eq:tripg}) that for small 
values of $\Delta_{i}(k_{F})$ the integrand 
\[
f(k,k')_{i}\equiv -\frac{1}{\pi}k'^{2}\left\langle k|V(^{3}P_{2})|k'
\right\rangle\frac{\Delta_{i}(k')}{\tilde{E}_{i}(k')}\;\;i=0,2 
\]
in the gap equation has a very narrow peak at $k=k_{F}$, as shown 
in fig. \ref{fig:fig2}.   
Thus, to obtain stable results, we had to take great care in 
choosing mesh points 
for the momentum space integration  near $k_{F}$.  As seen 
from fig. \ref{fig:fig2} the choice of upper cutoff momentum 
was almost immaterial, and $k=15\;{\rm fm}^{-1}$ was found to be adequate.  
We should mention that Khodel, Khodel and Clark \cite{kkc96} recently 
published a new method for solving gap equations which eliminates 
the problems occuring when the gap parameter is small.

\begin{figure}[htbp]
\begin{center}
{\centering
\mbox
{\psfig{figure=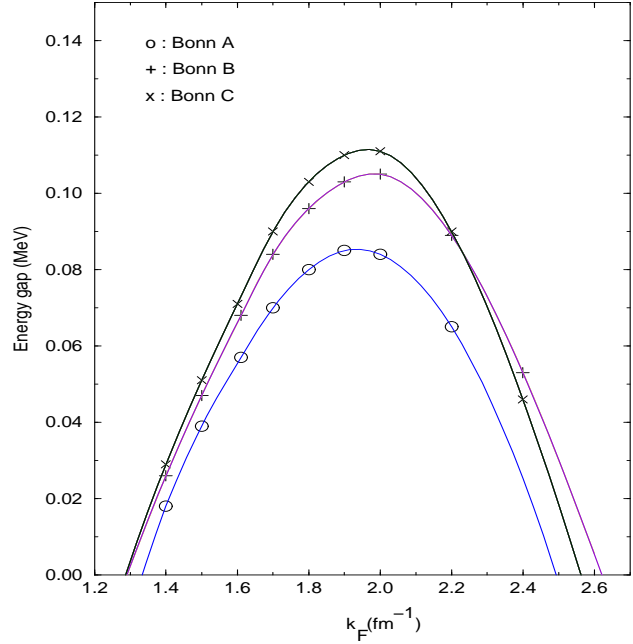,height=8cm,width=8cm}}
}
\caption{Angle-averaged energy gap $\overline{D}_{0}(k_{F})$ calculated 
with Bonn A, B and C.  
}
\label{fig:fig3}
\end{center}
\end{figure}

\begin{figure}[htbp]
\begin{center}
{\centering
\mbox
{\psfig{figure=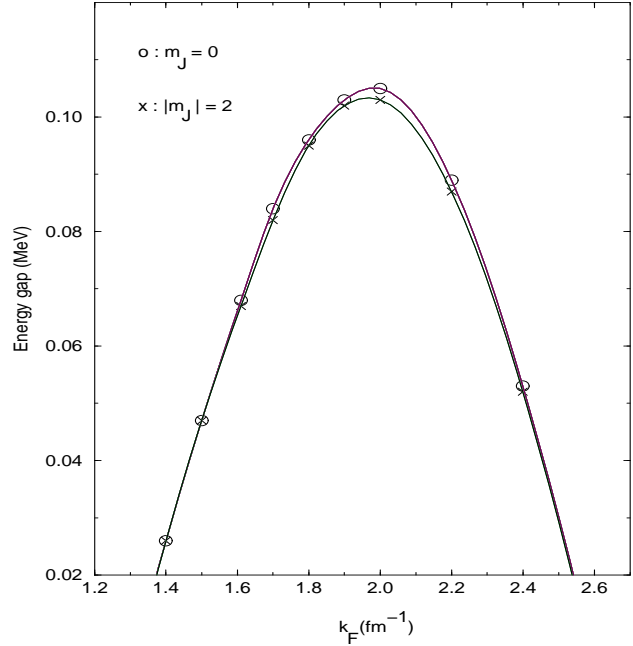,height=8cm,width=8cm}}
}
\caption{Comparison of the angle averaged energy gap for $m_{J}=0$ with 
corresponding results for $|m_{J}|=2$.  The results are for the 
Bonn B potential.  
}
\label{fig:fig4}
\end{center}
\end{figure}

In fig. \ref{fig:fig3} the energy gap $\overline{D}_{0}(k_{F})$, calculated 
with MBHF single-particle energies, is shown.    
The energy gap $\overline{D}_{2}(k_{F})$ was found to be similar to 
$\overline{D}_{0}(k_{F})$, as seen from fig. \ref{fig:fig4}.  
Takatsuka and Tamagaki \cite{tak93} and Amundsen and 
\O stgaard \cite{am85} found that the $^{3}P_{2}$ gap vanished for 
realistic values of the neutron effective mass (in the range 0.90-0.65, see 
table \ref{tab:tab1}), and thus it was 
necessary to include the attractive tensor coupling to the $^{3}F_{2}$ wave.
As will be shown in the next section, our results for the pure 
$^{3}P_{2}$ wave are of the same size as the results where 
the tensor coupling is included.  This result is probably due to the 
Bonn potentials having a weaker tensor force than the potentials used 
in refs. \cite{tak93,am85}.
However, our results are one order of magnitude smaller than the 
results of Baldo {\it et al.} \cite{bld92}, who found energy gaps  
of the order $1\;{\rm MeV}$ without 
the tensor coupling effect.    

The energy gap function $D_{0}$ given by eq. (\ref{eq:d0a}) is nodeless (as 
a function of $\theta$), while $D_{2}$ has nodes for $\cos^{2}\theta=1$.  
A nodeless energy gap results in the usual exponential decay of the 
specific heat below the superfluid transition temperature.  
If there are nodes, the specific heat will follow a power-law \cite{and61}.  
Thus the solution with $m_{J}=0$, if realized, gives a different thermal 
behaviour than the solution with $|m_{J}|=2$.  The favoured solution 
is the one which gives the lowest condensation energy.  This quantity 
is, in the so-called weak-coupling limit, given by 
\begin{equation}
\Delta E \approx -\frac{\rho_{0}}{2}\overline{D}^{2}_{\lambda}(k_{F}), 
\label{eq:conde}
\end{equation}
where $\rho_{0}$ is the density of states at the Fermi energy.  
\begin{table}[t]
\begin{center}
\begin{tabular}{ccc} 
\multicolumn{1}{c}{$k_{F}\;({\rm fm}^{-1})$}&\multicolumn{1}{c}{
$-\Delta E/(\rho_{0}/8\pi)$($m_{J}=0$) (MeV) }&\multicolumn{1}{c}
{$-\Delta E/(\rho_{0}/8\pi)$($|m_{J}|=2$) (MeV)} \\ \hline  
  1.40   &  0.008  &  0.008 \\ 
  1.60   &  0.058  &  0.056 \\
  1.80   &  0.113  &  0.116 \\
  2.00   &  0.139  &  0.133 \\ 
  2.20   &  0.100  &  0.095 \\ \hline 
\end{tabular}
\caption{Condensation energies calculated with the Bonn B potential.}
\label{tab:tab1}
\end{center}
\end{table} 
In table \ref{tab:tab1} the two solutions are compared with respect 
to the condensation energy.  The results shown are for the Bonn B potential.  
It is seen that the two solutions are very nearly degenerate, a result 
which is in agreement with the more complete discussions found in 
refs. \cite{tak93,am85}, with the nodeless solution, $m_{J}=0$, seemingly 
the favoured one.     

\section{Inclusion of the tensor coupling}

We have solved the coupled equations (\ref{eq:tgap2}) and (\ref{eq:tgap3}) 
for pure neutron matter.    
We give results for the angle average of $D_{\lambda}$,  
\begin{equation}
\overline{D}_{\lambda m_{J}=0}(k_{F})=\frac{1}{\sqrt{8\pi}}
\sqrt{\Delta_{\lambda 10}^{2}(k_{F})+\Delta_{\lambda 30}^{2}(k_{F})}
\approx \frac{\Delta_{\lambda 10}}{\sqrt{8\pi}}, 
\end{equation}
where the last approximation is valid because $\Delta_{\lambda 30}(k_{F})$ 
was found to be negligible in comparison with $\Delta_{\lambda 10}(k_{F})$.  
The following procedure was used to solve the coupled gap equations: 
First we obtained single-particle energies in the MBHF approach, as 
described in section 2.3.  Starting with suitable approximations for 
the gap components $\Delta_{\lambda 10}$ and $\Delta_{\lambda 30}$ 
(for $\Delta_{\lambda 10}$ we could start with the solution of 
the uncoupled equation considered in section 3), the gap equations 
(\ref{eq:tgap2}) and (\ref{eq:tgap3}) were solved by iteration.  Again, 
special attention had to be paid to the integration over momenta near 
$k_{F}$, because of the almost singular behaviour of the integrand there.  
Stable results were obtained with the same choice of mesh points as for 
the uncoupled case considered in the previous section.  

\begin{figure}[htbp]
\begin{center}
{\centering
\mbox
{\psfig{figure=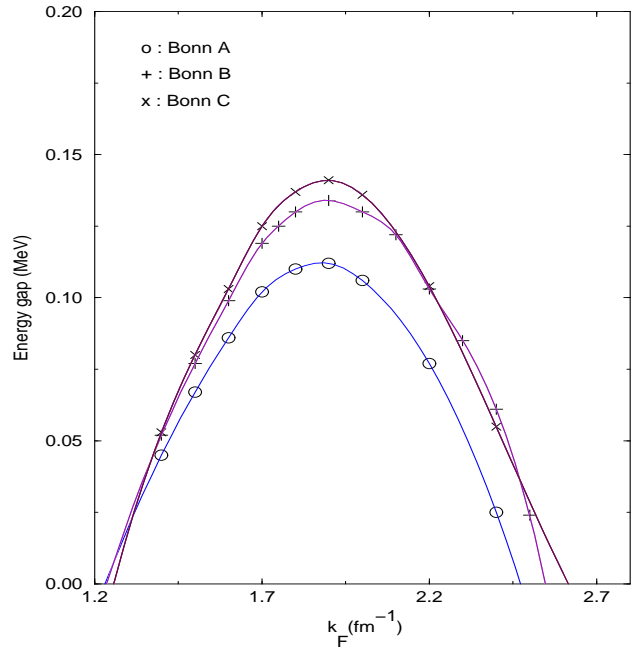,height=8cm,width=8cm}}
}
\caption{$^{3}P_{2}$-$^{3}F_{2}$ energy gaps in pure neutron matter.
}
\label{fig:fig5}
\end{center}
\end{figure}

\begin{figure}[htbp]
\begin{center}
{\centering
\mbox
{\psfig{figure=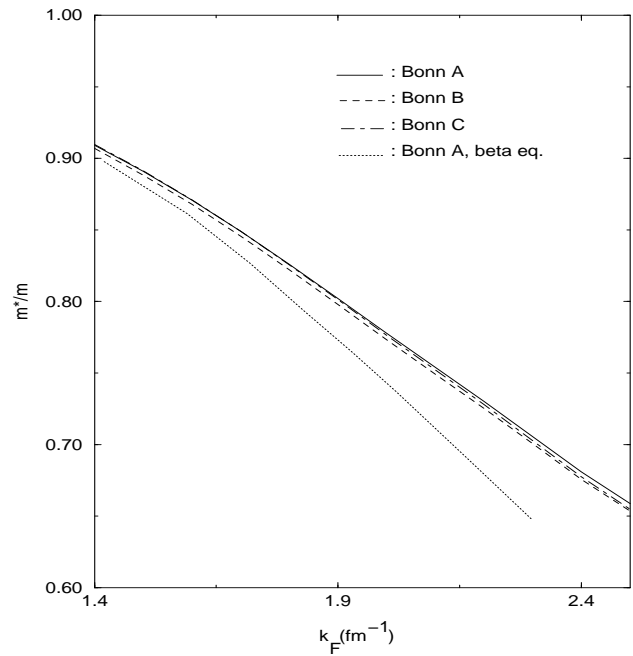,height=8cm,width=8cm}}
}
\caption{Effective masses in pure neutron matter and in matter at 
$\beta$ equilibrium.
}
\label{fig:fig6}
\end{center}
\end{figure}

In fig. \ref{fig:fig5} the energy gaps in pure neutron matter obtained 
with Bonn A, B and C are shown.  
From these results it is seen that the tensor force increases the 
energy gaps.  However, this effect is not so important here as in the 
works of Takatsuka and Tamagaki \cite{tak93} and Amundsen and \O stgaard 
\cite{am85}.  This was to be expected, since the Bonn potentials have 
weaker tensor components than the OPEG potentials used in refs. 
\cite{tak93,am85}.  The effects of the tensor coupling are greater 
for Bonn C than for Bonn B and A, reflecting the differing tensor 
strengths of the Bonn interactions.     
In fig. \ref{fig:fig6} the effective masses are plotted as functions 
of the neutron Fermi momentum.  This figure shows that the Bonn models 
give nearly identical effective masses in pure neutron matter.  
The differences in the energy gaps are therefore mainly due to 
differences in the  $^{3}P_{2}$-$^{3}F_{2}$ wave of the interactions.
From fig. \ref{fig:fig1} it is seen that Bonn C gives more attraction 
near $k_{F}$ than do Bonn B and A.  Since the main contribution to the 
integral in the gap equation comes from the region near $k_{F}$, it is 
reasonable that Bonn C gives the largest energy gaps.      

\begin{figure}[htbp]
\begin{center}
{\centering
\mbox
{\psfig{figure=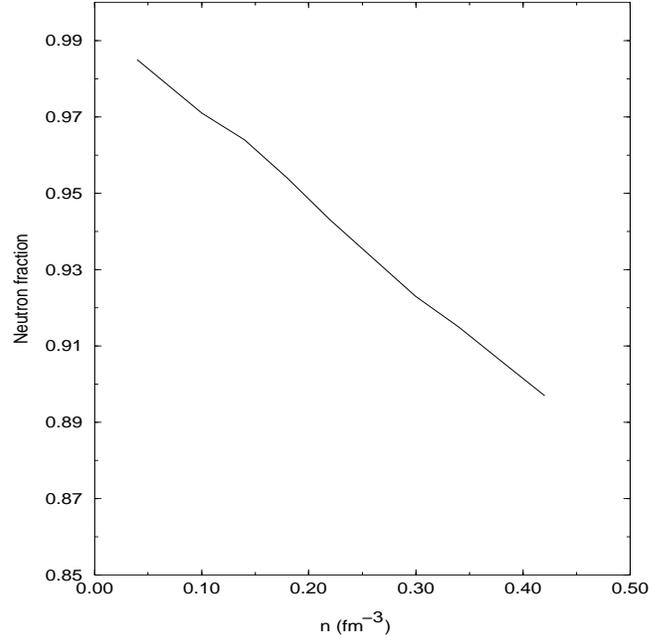,height=8cm,width=8cm}}
}
\caption{Neutron fraction as a function of the nucleon density $n$ at 
$\beta$ equilibrium with muons included.
}
\label{fig:fig7}
\end{center}
\end{figure}

\begin{figure}[htbp]
\begin{center}
{\centering
\mbox
{\psfig{figure=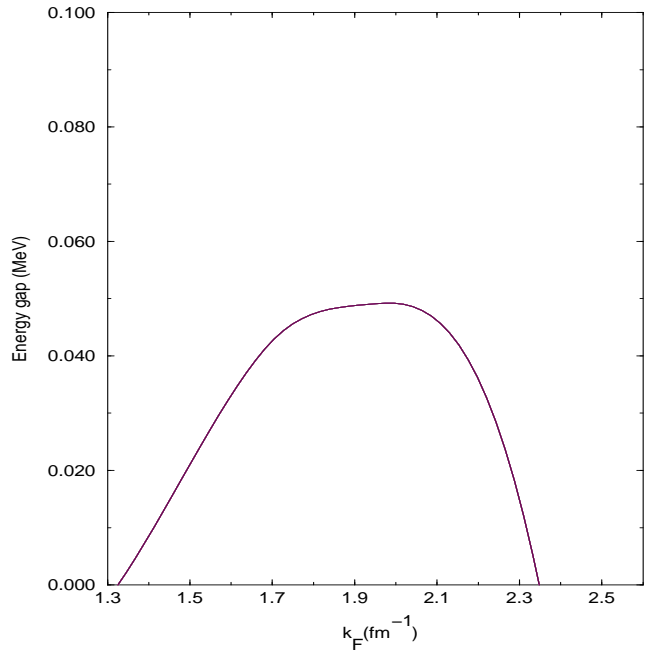,height=8cm,width=8cm}}
}
\caption{Neutron energy gap at $\beta$ equilibrium.
}
\label{fig:fig8}
\end{center}
\end{figure}

We have also calculated the $^{3}P_{2}$-$^{3}F_{2}$ neutron energy 
gap in neutron star matter at $\beta$ equilibrium.  The neutron 
fractions, shown in figure \ref{fig:fig7}, were determined from a 
self-consistent MBHF calculation, 
imposing the relevant equilibrium constraints as described in 
section 2.3.  The results, obtained with the Bonn A potential, are shown 
in figure \ref{fig:fig8}.
The size of the energy gap is reduced considerably, and the density 
range for this superfluid shrinks in comparison with the 
case of pure neutron matter.   The main reason for this is probably 
the reduced neutron effective masses in $\beta$-stable matter, as 
seen in fig. \ref{fig:fig6}.  

\section{Discussion}

\begin{figure}[htbp]
\begin{center}
{\centering
\mbox
{\psfig{figure=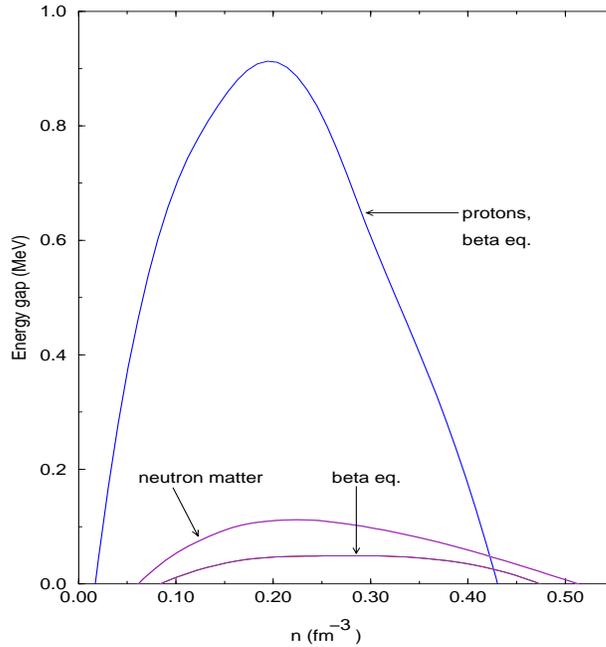,height=8cm,width=8cm}}
}
\caption{Energy gaps for protons and neutrons in the quantum fluid 
region of a neutron star.
}
\label{fig:fig9}
\end{center}
\end{figure}

The results of this paper are summarized in fig. \ref{fig:fig9}, where 
we show the following results:
\begin{itemize}
 \item the $^{3}P_{2}$-$^{3}F_{2}$ energy gap in pure neutron matter
 \item neutron $^{3}P_{2}$-$^{3}F_{2}$ energy gap at $\beta$ equilibrium
 \item proton $^{1}S_{0}$ energy gap at $\beta$ equilibrium, taken from 
       ref. \cite{elg96}, 
\end{itemize}
all as functions of the nucleon density.  From 
the figure it is seen that the proton energy gap is one order of magnitude 
larger than the neutron energy gaps.  Thus the dominating contribution 
to the suppression of the modified URCA processes will be due to 
superconducting protons \cite{elg95}.   
Taking into account that in a neutron star we have matter at 
$\beta$ equilibrium and not pure neutron matter, the neutron energy 
gap is reduced by approximately a factor of two.  

\begin{figure}[htbp]
\begin{center}
{\centering
\mbox
{\psfig{figure=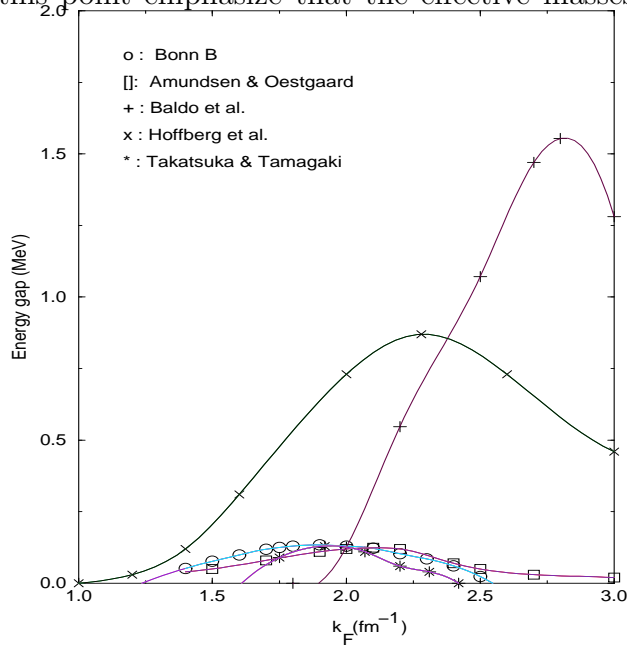,height=8cm,width=8cm}}
}
\caption{Results for the triplet energy gap in neutron matter.  
The results of Baldo {\it et al.} and Hoffberg {\it et al.} were obtained 
without the tensor coupling.
}
\label{fig:fig10}
\end{center}
\end{figure}

\begin{figure}[htbp]
\begin{center}
{\centering
\mbox
{\psfig{figure=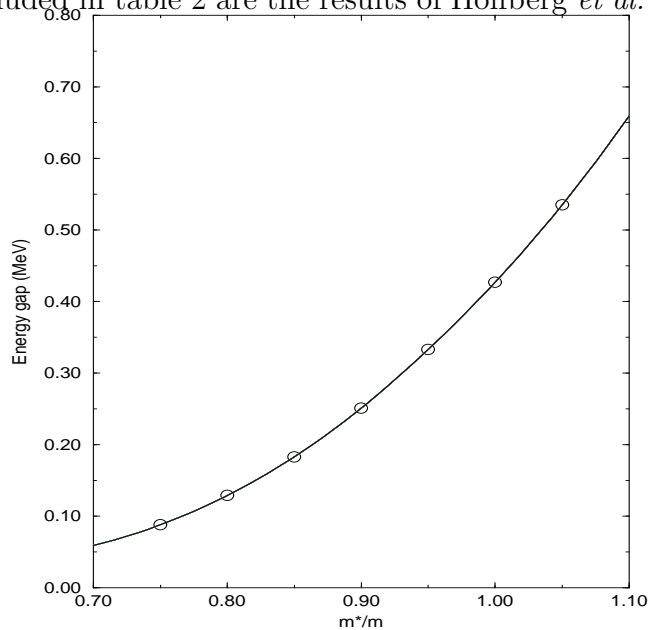,height=8cm,width=8cm}}
}
\caption{Effective mass dependency of the $^{3}P_{2}$-$^{3}F_{2}$ energy 
gap.  The results were obtained using the Bonn C potential and 
$k_{F}=1.8\;{\rm fm}^{-1}$.
}
\label{fig:fig11}
\end{center}
\end{figure}

The results obtained in this work for the $^{3}P_{2}$-$^{3}F_{2}$ energy 
gaps in pure neutron matter are in good agreement with those of 
Takatsuka and Tamagaki \cite{tak93} and Amundsen and \O stgaard \cite {am85}.
We find a maximum energy gap of the same size as in refs. \cite{tak93,am85}, 
and a density region for pairing intermediate between those found in the 
two references above.  
In table \ref{tab:tab2} and fig. \ref{fig:fig10} we summarize  
the results of refs. \cite{tak93}(TT), \cite{bld92}(BCLL), 
\cite{am85}(A\O), \cite{hof70}(HGRR), and the present work.
\begin{table}[t]
\begin{center}
\begin{tabular}{lllll} 
\multicolumn{1}{c}{Authors}&\multicolumn{1}{c}{Max. energy gap (MeV)}&
\multicolumn{1}{c}{Density range (${\rm fm}^{-1}$)}& 
\multicolumn{1}{c}{$m^{*}/m$ (typical values)} &
\multicolumn{1}{c}{Potential} \\ \hline
  A\O    &  0.12   &  1.4-3.0 &  0.70-0.89 & OPEG \\
  BCLL   &  1.5    &  1.8-3.2 &  0.70-0.86 & Argonne $V_{14}$ \\
  TT     &  0.12   &  1.5-2.4 &  0.70-0.82 & OPEG  \\  
  HGRR   &  0.9    &  1.0-3.0 &  1.0       & Tabakin \\
This work&  0.13   &  1.2-2.5 &  0.65-0.91 & Bonn B   \\ \hline
\end{tabular}
\caption{Results for the $^{3}P_{2}$ energy gap in neutron matter.}
\label{tab:tab2}
\end{center}
\end{table} 
We should at this point emphasize that the effective masses quoted in 
the table are meant to facilitate comparison between the different works.  
Both in this work and in refs. \cite{tak93,bld92}, the gap equations 
were solved with  $k$-dependent single-particle potentials, without 
resorting to the effective mass approximation.  
From table \ref{tab:tab2} it is seen that the the energy gap has generally  
been found to be of order $0.1\;{\rm MeV}$.  The results of Baldo {\it et al.} 
\cite{bld92} are remarkable, considering that they have 
neglected the attractive contributions from the tensor coupling to 
$^{3}F_{2}$.  With effective masses in the range used by Baldo {\it et. al}, 
the energy gap in the pure $^{3}P_{2}$ wave was in this work found to be one 
order of magnitude smaller, and in refs. \cite{tak93,am85} it was found to 
vanish.  In addition, we found the $^{3}P_{2}$ gap to exist at lower 
densities than Baldo {\it et al}.     
Although both Amundsen and \O stgaard  and Takatsuka and Tamagaki 
used the OPEG potentials, Amundsen and \O stgaard found that the 
energy gap extended up to $k_{F}=3.0\;{\rm fm}^{-1}$.  This difference 
can be understood because different approaches to the single-particle 
energies were employed.  Amundsen and \O stgaard took $m^{*}/m$ to 
be constant (equal to 
0.70) above $k_{F}=2.4\;{\rm fm}^{-1}$ \cite{amth83}, while Takatsuka and 
Tamagaki used 
a density-dependent effective mass all the way, and thus it dropped to 
values below $0.70$ beyond $k_{F}=2.4\;{\rm fm}^{-1}$.  Thus it is 
seen that the single-particle spectrum is a crucial ingredient in 
energy gap calculations.  However, the Brueckner-Hartree-Fock spectrum,  
and variations thereof, used in most works, including the present, may be 
of questionable validity at the densities where the $^{3}P_{2}$ energy 
gap closes, where relativistic effects and many-body correlations become 
increasingly important.      
To further illustrate the importance of the single-particle energies, 
we have calculated the energy gap at $k_{F}=1.8\;{\rm fm}^{-1}$ as 
a function of the effective mass parameter $m^{*}/m$.  The   
results in fig. \ref{fig:fig11} 
show that the energy gap depends strongly on the effective mass.    
Also included in table \ref{tab:tab2} are the results of Hoffberg 
{\it et al.} \cite{hof70}.  
They are, however, not directly comparable with 
the other results, as they have been obtained directly from the 
$^{3}P_{2}$ phase shifts, and with $m^{*}/m=1$.  
However, it is interesting to note that they obtained a smaller maximum 
energy gap than Baldo {\it et al}.   

To the best of our knowledge, the $^{3}P_{2}$-$^{3}F_{2}$ energy gap 
has not been calculated in $\beta$-stable neutron star matter before.  
Our results show that the energy gap is reduced by a factor of two when 
$\beta$ equilibrium is taken into account, mainly because the neutron 
effective mass is reduced.  A point of interest here is that in a 
recent investigation of neutron star cooling, Schaab {\it et al.} 
\cite{scha96} 
found that agreement with observations was obtained if the $^{3}P_{2}$ 
energy gaps were lowered by a factor of two (they used the results 
of ref. \cite{am85} for pure neutron matter).  
A recent 
analysis by Page \cite{pag94} of data for the Geminga pulsar 
showed that baryon pairing in most of the star was necessary to 
obtain agreement with the observed surface temperature, assuming that 
fast cooling processes were possible.  However, with equations of 
state for $\beta$-stable matter, calculated within the non-relativistic 
BHF approach, we find that the critical density for the onset of the 
direct URCA processes to be $0.62\;{\rm fm}^{-3}$, while within the 
relativistic Dirac-Brueckner-Hartree-Fock approach, the corresponding 
density was found to be 
$0.50\;{\rm fm}^{-3}$ \cite{elg95}.  At these densities the neutron and 
proton pairing gaps have already vanished, so the direct URCA processes 
will not be suppressed by superfluidity within this picture.     
Further investigation of $^{3}P_{2}$ pairing is therefore of interest.
However, an important shortcoming of all calculations, 
including those reported 
in this work, is the neglect of many-body effects in the pairing 
interaction.  The so-called polarization term has been found 
to reduce the $^{1}S_{0}$ proton and neutron energy gaps by as much 
as a factor of 2-3 \cite{ain89}.  It is, however, expected 
that polarization effects will tend to increase the $^{3}P_{2}$ neutron 
energy gap \cite{peth91}.  Further investigation of superfluidity in dense 
matter is therefore of interest.    

\section{Conclusion}

In this work we have calculated pairing gaps for neutrons in the 
$^{3}P_{2}$ state in 
neutron matter and in matter at $\beta$ equilibrium.  For 
pure neutron matter, we found a non-zero energy gap for 
$1.2\;{\rm fm}^{-1}\leq k_{F} \leq 2.5\;{\rm fm}^{-1}$, with a maximum 
value of $0.12-0.13\;{\rm MeV}$.  Taking the conditions 
for $\beta$ equilibrium 
into account, the corresponding values became $1.32\;{\rm fm}^{-1}
\leq k_{F} \leq 2.34\;{\rm fm}^{-1}$ and $0.05\;{\rm MeV}$.  
However, evaluation of the contribution from medium polarization effects 
to the pairing interaction is necessary before any definite conclusions 
can be drawn.

This work has been supported by the Research Council of Norway (NFR) 
under the Programme for Supercomputing through a grant of computing time.  
MHJ thanks the NFR and the Istituto Trentino di Cultura, Italy, 
for financial support.

\end{document}